\documentstyle[preprint,aps,epsf]{revtex}

\newcommand{\la}{\langle}
\newcommand{\ra}{\rangle}
\newcommand{\beq}{\begin{eqnarray}}
\newcommand{\eeq}{\end{eqnarray}}

\newcommand{\btem}{\bibitem}

\tighten

\begin{document}

\preprint{UTHEP-363, June 1997}

\draft

\title{Soft Modes associated with Chiral Transition
at Finite Temperature}

\author{S. Chiku\cite{JSPS} and T. Hatsuda}

\address{Institute of Physics, University of Tsukuba,
 Tsukuba, Ibaraki 305, Japan}

\date{\today}
 
\maketitle

\begin{abstract}
 Using a novel resummation procedure of thermal loops,
 real-time correlations in the scalar and pseudo-scalar channels
 are studied in the $O(4)$ linear
 $\sigma$ model at finite temperature. 
 A threshold enhancement of the spectral function in the
 scalar channel is shown to be a noticeable  precritical phenomenon
 of the chiral phase transition.

\end{abstract}

\narrowtext

\noindent
  Observing the restoration of chiral symmetry at 
 finite-temperature ($T$) is one of the 
 central aims in the future relativistic heavy-ion experiments
 at RHIC and LHC \cite{QM96}.
 Numerical studies of  quantum chromodynamics (QCD) on the lattice
 for two and three flavors are actively pursued to reveal the 
 nature of the chiral transition \cite{lattice}.

 From the theoretical point of view,
  one of the ideal signals of  the second-order (or
 weakly first order) chiral transition 
 is the existence of the soft modes associated with
  chiral order parameters  
 $\phi({\bf x},t) = (\bar{q}q ({\bf x},t),
 \bar{q}i \gamma_5 \vec{\tau } q ({\bf x},t)) $ \cite{shift}.
 The softning  is a 
 {\em dynamical} phenomena
  characterized by the anomalous enhancement of 
 the dynamic susceptibility defined by
\beq
 D^R_{\phi}(\omega, {\bf k}) =  -i  \int d^4x e^{i kx}
  \theta(t) \la  [\phi({\bf x},t) \phi(0,0)] \ra_{_T} \  .
\eeq 
 This phenomena at finite $T$ was first
 discussed in ref.\cite{HK} using an effective theory of QCD.
 Also there are subsequent theoretical and 
  phenomenological studies  on this problem \cite{HULI}.
 In condensed matter and solid state physics,  
 the soft modes have been studied extensively 
 in neutron and light scattering experiments \cite{GK}.

 The purposes of this paper are twofold. First one  is  
 to study the spectrums of the soft modes at finite $T$
 by taking into account the strong coupling between 
 the scalar and pseudo-scalar channels.
 Despite the fact 
 that this has a considerable effect on the  qualitative
 behavior of $D^R_{\phi}$, it has not been studied seriously
  so far.
  The second purpose is to present a
 systematic resummation procedure of thermal loops
 which is  applicable even when  the dynamical symmetry breaking takes place.
 We will show that 
 this procedure is inevitable for studing the first problem.

 To demonstrate  the above points,
 we adopt the  $O(4)$ linear $\sigma$ model
 with explicit symmetry breaking.
 The model  has been used
 to study  critical exponents of the chiral transition \cite{PW}
 on the basis of the static and dynamical universality. 
 The Lagrangian density reads
\beq
\label{lag1}
{\cal L} & = & {1 \over 2} [(\partial_{\mu} \phi_i )^2
 + \mu^2 \phi_i^2 ] -{ \lambda \over 4! } (\phi_i^2)^2 + h \phi_0 \\ \nonumber
  & &  +\  {\rm counter}\ \ {\rm terms},
\eeq
with $\phi_i = (\sigma, \vec{\pi})$,  and $h \phi_0$ being the explicit
 symmetry breaking term. 
 All the divergences are removed by  counter terms 
  in the symmetric phase ($\mu^2 < 0$)
 in the $\overline{MS}$ scheme.
 In this symmetric and  mass independent scheme,
  the renormalization
 constants are $\mu^2$ independent  \cite{BWL}.  
  The lagrangian after dynamical symmetry breaking ($\mu^2 > 0$)  is  
 obtained from (\ref{lag1}) by the replacement 
 $\sigma \rightarrow \sigma + \xi$, where  
  $\xi$ is determined by  the stationary condition for the
 effective potential $\partial V(\xi)/\partial \xi =0$.

 Causal meson propagators at $T=0$ have a general form 
 $D_{\phi}(q) = [q^2 -m_{0\phi}^2 - \Sigma_{\phi} (q^2)
 + i \epsilon ]^{-1}$, where 
 $m_{0\phi}$ is the tree-level mass, namely  
 $m_{0\sigma}^2  = - \mu^2 + \lambda \xi^2 /2$
 and  $m_{0\pi}^2  = - \mu^2 + \lambda \xi^2 /6$.
 The one-loop calculation of the self-energy 
  $\Sigma_{\phi}$ 
  is standard \cite{BWL} and we do not  recapitulate it here.

 The renormalized couplings 
 $\mu^2$, $\lambda$ and $h$ are 
 determined by the following renormalization conditions:
 (i) on-shell condition for pion, 
   $D^{-1}_{\pi} (q^2 = m_{\pi}^2) = 0$ with $m_{\pi}=140$ MeV,
 (ii) PCAC in one-loop,  
 $ f_{\pi} m_{\pi}^2 = h \sqrt{Z_{\pi}}$ with
 $Z_{\pi}$ being the wave function renormalization constant for
 pion and $f_{\pi} = 93 $ MeV,
 (iii) the peak position of the spectral function 
 $\rho_{\sigma}= -(1/\pi){\rm Im}D_{\sigma}$ is chosen to be 550 MeV;
  this number corresponds to the value obtained in the recent 
 re-analyses of the $\pi$-$\pi$ scattering phase shift \cite{pipi}.
  We have also taken 750 MeV  and 1 GeV \cite{HK}, and checked that
  our main conclusions  do not suffer qualitative change.
  The arbitrary renormalization point
 $\kappa$ is chosen so that
$\rho_{\sigma}(\omega)$ starts from 
  the correct continuum threshold at $\omega = 2 m_{\pi}$:
 This is achieved by demanding $m_{0\pi}$=$m_{\pi}$.
 The resultant values read $ \mu^2 = (283 \ {\rm MeV})^2$,
 $\lambda = 73.0$, $h = (123 \ \mbox{MeV})^3$ and
 $\kappa = 255 \ \mbox{MeV}$.

 As has been known
 since the works of  Weinberg, and Kirzhnits and Linde \cite{WEIN},
  naive perturbation theory breaks
 down for sufficiently high $T$.
 In the linear $\sigma$ model,
 this is easily seen from the behavior of $m_{0\phi}$ defined above
  which becomes tachyonic as one approaches to the 
  symmetric phase
  ($\xi \rightarrow 0$). This not only destroys the credibility of the
 loop expansion at high $T$, but also leads to an 
  unphysical threshold in $\rho_{\sigma}(\omega)$ 
  even at low $T$. The latter aspect is particularly 
 harmful for the purpose of this paper. 

  To cure these problems,  we developed a method 
 based on  a  resummation procedure
  proposed by  Banerjee and Mallik \cite{BM}.
 It  is a generalized  mean-field theory
 and  allows one to
 carry out  systematic perturbation theory and renormalization 
 at finite $T$ \cite{HF}.
  In  high $T$ limit, the method reduces to 
 the resummation of hard thermal loops \cite{BP}.
 We will report the technical details elsewhere \cite{CH}, and
 recapitulate here only the essential parts needed for  
 subsequent discussions.

 Let us rewrite eq.(\ref{lag1}) by adding and subtracting
  a chiral invariant mass
 term with an arbitrary  mass  $m$;
\beq
\label{lag2}
{\cal L} & = & {1 \over 2} [(\partial_{\mu} \phi_i )^2
 - m^2 \phi_i^2] -{ \lambda \over 4! } (\phi_i^2)^2 + h \phi_0 \\ \nonumber
  & &  + {1 \over 2} (A-1) (\partial_{\mu} \phi_i )^2
       - {1 \over 2} B m^2 \phi_i ^2 
     - { \lambda \over 4! } C (\phi_i^2)^2  \\ 
  & &   + {1 \over 2} (m^2 + \mu^2)(1+B) \phi_i^2
      + h(\sqrt{A}-1) \phi_0 \ \ \ .  \nonumber
\eeq
$A$, $B$ and $C$ are renormalization constants:
 In one-loop in the $\overline{MS}$ scheme,
 $A=1$, $B = \lambda /16 \pi^2 \bar{\epsilon}$
 and $C = \lambda /8 \pi^2 \bar{\epsilon}$.
 The loop expansion at finite $T$
 can be done in the same way as that at $T=0$  except that
 (i) $m^2$ should be used instead of $-\mu^2$, and (ii)
 a new vertex proportional to $m^2 + \mu^2$ appears.
 $m^2$ is a $T$ dependent mass parameter to be determined later 
 by  the dynamics.

At finite $T$, retarded meson propagators read
\beq
D^R_{\phi}(\omega,{\bf k})
  = [k^2 - m_{0\phi}^2(T) - \Sigma^R_{\phi}
 (\omega, {\bf k};T)]^{-1},
\eeq
where $k^2 = \omega^2 - {\bf k}^2 $ and 
\beq
 m_{0 \pi}^2(T)  =  m^2 + {\lambda \xi^2  \over 2},
 \ \ \  m_{0 \sigma }^2(T) & = & m^2 + {\lambda \xi^2 \over 6} .
\eeq
 The self-energy 
   $\Sigma^R_{\phi}$ is obtained either from the imaginary-time
 or the real-time formalism. We adopt the latter in which
 ${\rm Re} \Sigma^R_{\phi} (\omega, {\bf k};T)
 = {\rm Re} \{\Sigma^{11}_{\phi}(\omega,{\bf k})+\Sigma^{11}_{\phi}
 (\omega, {\bf k};T)\}$ and ${\rm Im} \Sigma^R_{\phi} (\omega, {\bf k};T)
 = {\rm tanh}(\omega /2T)\  {\rm Im} \{\Sigma^{11}_{\phi}(\omega,{\bf k})+
 \Sigma^{11}_{\phi} (\omega, {\bf k};T)\}$.
 Here $\Sigma^{11}_{\phi}(\omega,{\bf k})$ ($\Sigma^{11}_{\phi}
 (\omega, {\bf k};T))$ is a $T$-independent ($T$-dependent)
 part of the 11-component of the 2 $\times$ 2 
  self-energy in the real-time formalism
 \cite{real}. Associated
 diagrams in one-loop with our modified  loop-expansion are shown in Fig. 1. 
 
 $m^2$ is a fictitious parameter and physical quatities should not
 depend on it. This leads us to several procedures
  to choose optimal $m^2$ in
 perturbation theory,
  such as the principle of minimal sensitivity (PMS), the    
 fastest apparent convergence (FAC) and so on \cite{pms}.
 We find that a variance of FAC is most suited for the purpose  
 of this paper: A condition for making the loop-correction to the
 real part of the pion propagator  as small as possible reads
\beq
\omega^2-m^2_{0\pi}- {\rm Re} [ \Sigma^{11}_{\pi}(\omega,0)+
\Sigma^{11}_{\pi}(0,0;T)] \ \vert_{\omega=m_{0\pi}} = 0.
\eeq
 $\omega $ in $ \Sigma^{11}_{\pi}(\omega,0;T)$ 
 is chosen to be zero by a technical reason 
 for getting a continuous solution of $m^2$ as a function of $T$
  \cite{CH}.
  Our procedure naturally gives
 $m^2 \rightarrow - \mu^2$ as $T \rightarrow 0$
 and  $m^2 \rightarrow \lambda T^2/12 $ as $T \rightarrow \infty$.
 The former guarantees that the loop-expansion at $T=0$ is
 not spoiled, and the latter guarantees that
   thermal tadpole diagrams are resummed correctly.

 In Fig.2(A), the chiral condensate
 $\xi (T)$ obtained by  minimizing the effective potential
 is shown for $m_{\pi} (T=0) =140$ MeV as well as for
  $m_{\pi} (T=0) = 50 $MeV. The latter corresponds to 
 a fairly small quark mass compared to the physical value 
 $m_q^{\rm phys.}$:
 $m_q / m_q^{\rm phys.} \simeq 
 (50 {\rm MeV}/140 {\rm MeV})^2 = 0.13$ \cite{order}.

In Fig.2(B), 
 $m_{0 \pi}(T)$, $m_{0 \sigma }(T)$ and  $m^2(T)$
 are shown.
 Because of our resummation procedure, 
 {\em tree-level masses}
  $m_{0 \pi}(T)$ and $m_{0 \sigma }(T)$ do not show tachyonic behavior 
 and both approach to the classical plasma limit ${\lambda T^2 /12}$
 at high $T$.
 This behavior is  important to have physical continuum threshold 
 in the spectral functions, bacause the threshold is dictated by 
 the particle masses running inside the loops (see Fig.1).

Let us now turn to the discussion on the spectral functions
 (imaginary part of the dynamical susceptibility) defined 
 by $\rho_{\phi} = -(1/\pi) {\rm Im}D^R_{\phi}$;
\beq
\label{spect2}
\rho_{\phi}(\omega,{\bf k})
  =  - {1 \over \pi}\ {{\rm Im} \Sigma^R_{\phi} \over
       (k^2 - m_{0\phi}^2 - {\rm Re}\Sigma^R_{\phi})^2 +
  ({\rm Im}\Sigma^R_{\phi} )^2 }.
\eeq
 In Fig.3(A),(B), $\rho_{\phi}(\omega, {\bf k}=0)$ is shown in 
 $\pi$ and  $\sigma$ channels  for $T=0,120,145$ MeV.
  In the $\pi$ channel at $T=0$, there is a distinct pion pole as well
  as a continuum starting from $m_{0 \pi} + m_{0 \sigma} \simeq 690 $MeV.
 In the  $\sigma$ channel at $T=0$, continuum starts 
 at $2 m_{0 \pi} = 280 $MeV. Also, there is a 
 broad peak centered at $\omega $= 550 MeV with a total width of 
 260 MeV.
 This corresponds to a $\sigma$-pole located far from the real axis
 in the complex $\omega$ plane. The large width is due to a strong 
  coupling of $\sigma$ with $2\pi$ in the linear $\sigma$ model.
 If we choose parameters so that the peak position of $\sigma$ are
 750 MeV (1 GeV), corresponding width reads 657 MeV (995 MeV).

 In  the $\pi$-channel for $ T \neq 0 $,
  a continuum arises in $ 0 < \omega < m_{0\sigma}-m_{0\pi}$.
  This is caused by the induced ``decay'' through the process
 $\pi + \pi \rightarrow \sigma $. Besides this,
 the pion still has its quasi-particle feature with no
 appreciable modification of the mass. This is in accordance with
 the Nambu-Goldstone nature of the pion.

 In  the $\sigma$-channel for $0 < T < 145 {\rm MeV}$, 
 there are two noticeable modification of the spectral function.
 One is the shift of the $\sigma$-peak toward the low mass region.
   The other is the sharpening of the
 spectral function just above the continuum threshold starting from
  $\omega = 2 m_{0\pi}(T)$.

  These features are actually related with each other and can 
 be understood in the following way.
  Because of the partial restoration of chiral symmetry 
 at finite $T$ together with the strong $\sigma$-$2\pi$ coupling,
 the real part of $(D_{\sigma}^R)^{-1}$ (which appears in  
 the first term of the denominator of eq.(\ref{spect2})) approaches
 zero at $\omega = 2 m_{0\pi}$ for $T \sim 145$ MeV 
 as shown in Fig.3(C). (The cusp structure
 in this figure originates from the pion-loop contribution to
   ${\rm Re} \Sigma^R_{\sigma}$.)
 In this situation,
  $\rho_{\sigma}(\omega \sim 2 m_{0\pi}) \sim
 1/{\rm Im}\Sigma^R_{\sigma} \propto
  \theta(\omega - 2m_{0\pi}) /\sqrt{1-(2 m_{0\pi}/\omega)^2 }$,  
 which shows a  singular behavior just above the threshold.

 In other words, the threshold enhancement in Fig.3(B), 
 although it occurs
 at relatively low $T$, is caused by a combined effect of the
 partial restoration of chiral symmetry and the strong $\sigma$-$2\pi$
 coupling. Note also that,
  near the chiral limit, 
the continuum threshold $2m_{0\pi}$ approaches to zero
 and the threshold enhancement occurs at the critical temperature
 where the chiral transition takes place.

 The spectral functions of $\pi$ ($\sigma$)
  for $T > 165 \ (145) $ MeV
 behave in a standard way as expected from the
 previous analyses \cite{HK,HULI}: 
  Simple $\sigma$ and $\pi$ poles without width 
 appear, because the decay ($\sigma \rightarrow 2 \pi$) and induced decay
 ($\pi + \pi \rightarrow \sigma$) are forbidden kinematically.
 As $T$ increases, these simple poles 
  gradually merge into a degenerate (chiral symmetric) state.
 For sufficiently 
  high $T$,
  the system is supposed to be in the deconfined phase and the decay
 $(\sigma, \pi) \rightarrow q \bar{q}$ starts to occur. This 
 is not taken into account in the present linear $\sigma$ model.
 A calculation based on the Nambu-Jona-Lasinio model shows, however, that
 there is still a chance for collective modes to survive
 as far as  $T/T_c$ is not so far from unity \cite{HK}.

In summary,
 we have studied the spectral function of the
 soft modes associated with chiral transition
 on the basis of  a special resummation method at finite $T$.
 An enhancement of the continuum threshold in the
 scalar channel are shown as a typical signal of the 
 partial restoration of chiral symmetry.
  Detectability of this phenomenon 
 in  experiments through the
 decays such as $\sigma \rightarrow 2 \pi, 2 \gamma, e^+e^-$
 remains as an interesting future problem.

\section*{Acknowledgements}

 The authors would like to thank T. Kunihiro for helpful discussions.
 This work was partially supported by
  the Grants-in-Aid of the Japanese Ministry of 
Education, Science and Culture (No. 06102004).

\onecolumn

\vspace{0.5cm}
\begin{figure}[h]
%\begin{center}
%\epsfile{file=FIG1.eps,scale=0.40}
%\end{center}
\caption{One-loop self-energy $\Sigma^{11}$ for $\sigma$ and $\pi$ in the
 modified loop-expansion at finite $T$.}
\label{fig:fig1}
\end{figure}

\begin{figure}[h]
%\begin{center}
%\epsfile{file=FIG2a.eps,scale=0.42}
%\epsfile{file=FIG2b.eps,scale=0.42}
%\end{center}
\caption{(A) $\xi(T)$ for $m_{\pi}=140 $MeV and 50 MeV.
(B) Masses in the tree-level $m_{0\pi}(T)$ and 
  $m_{0 \sigma}(T)$ shown with left vertical scale, and
 the mass parameter $m^2(T)$  with right vertical scale.}
\label{fig:fig2}
\end{figure}

\begin{figure}[h]
%\begin{center}
%\epsfile{file=FIG3ab.eps,scale=0.42}
%\epsfile{file=FIG3c.eps,scale=0.42}
%\end{center}
\caption{Spectral function in $\pi$ channel (A) and in $\sigma$ channel
 (B) for $T=0, 120, 145 $ MeV.
  The real part of $(D_{\sigma}^R(\omega, 0;T))^{-1}$
  as a function of $\omega$ is shown in (C).}
\label{fig:fig3}
\end{figure}

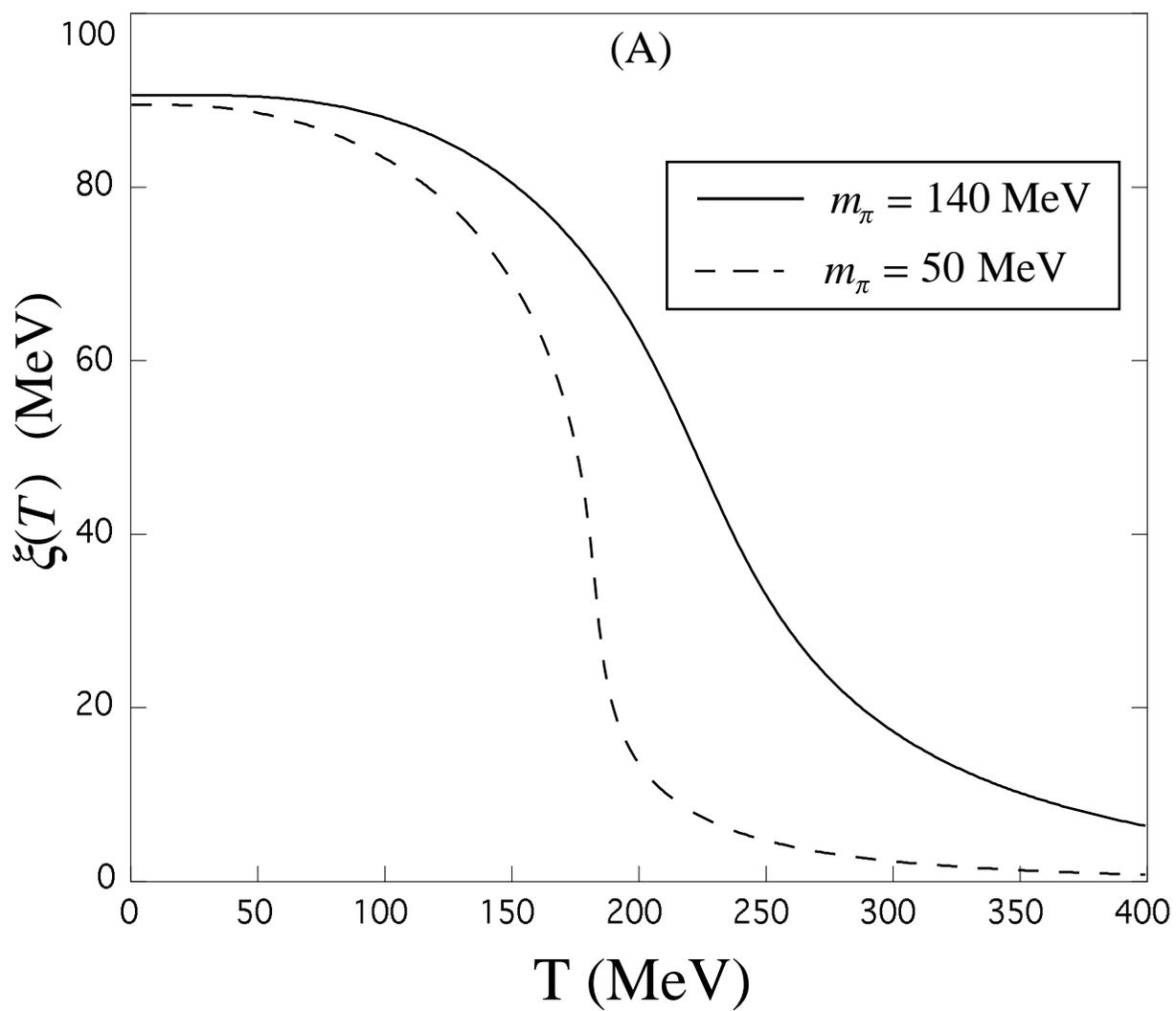
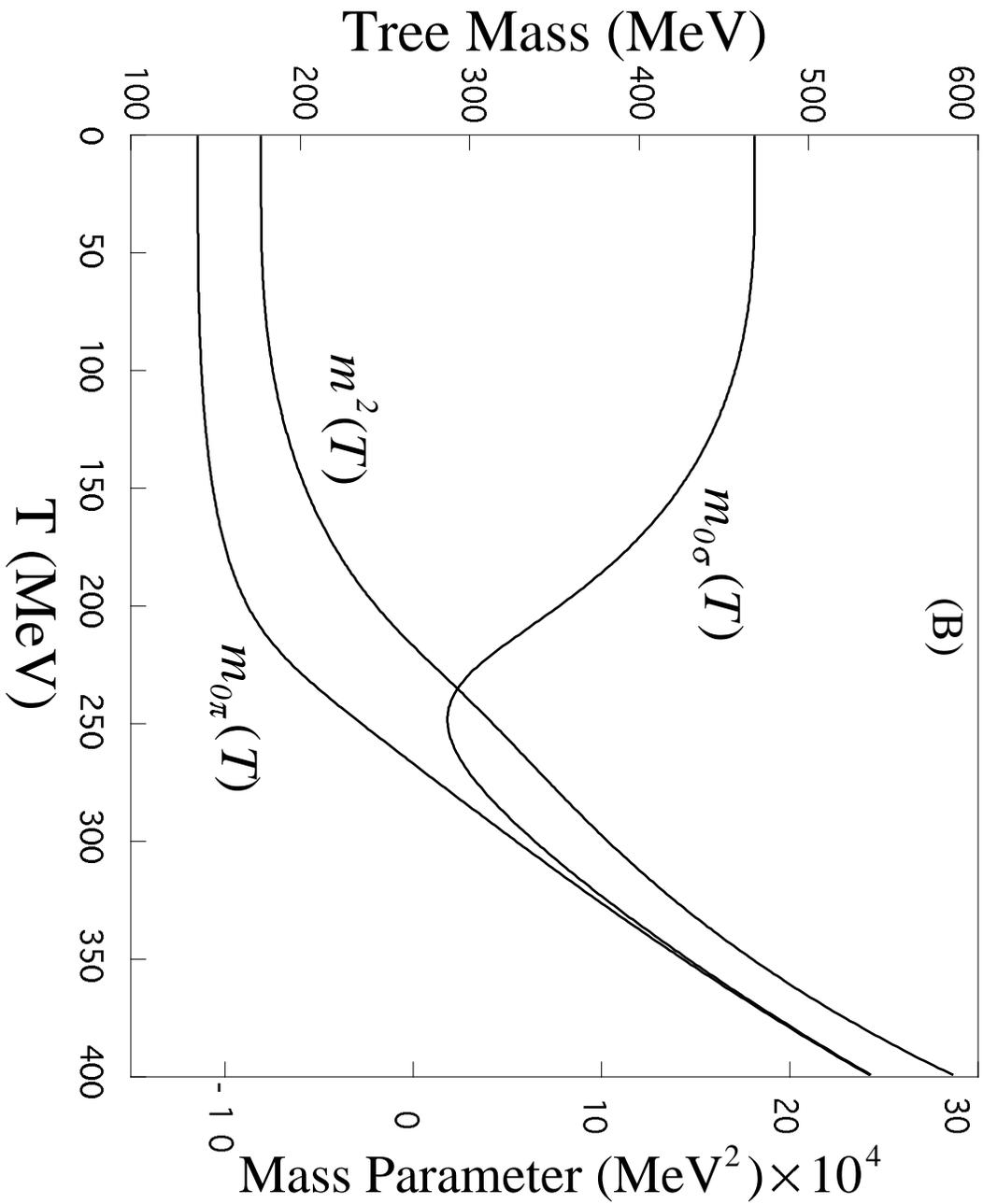
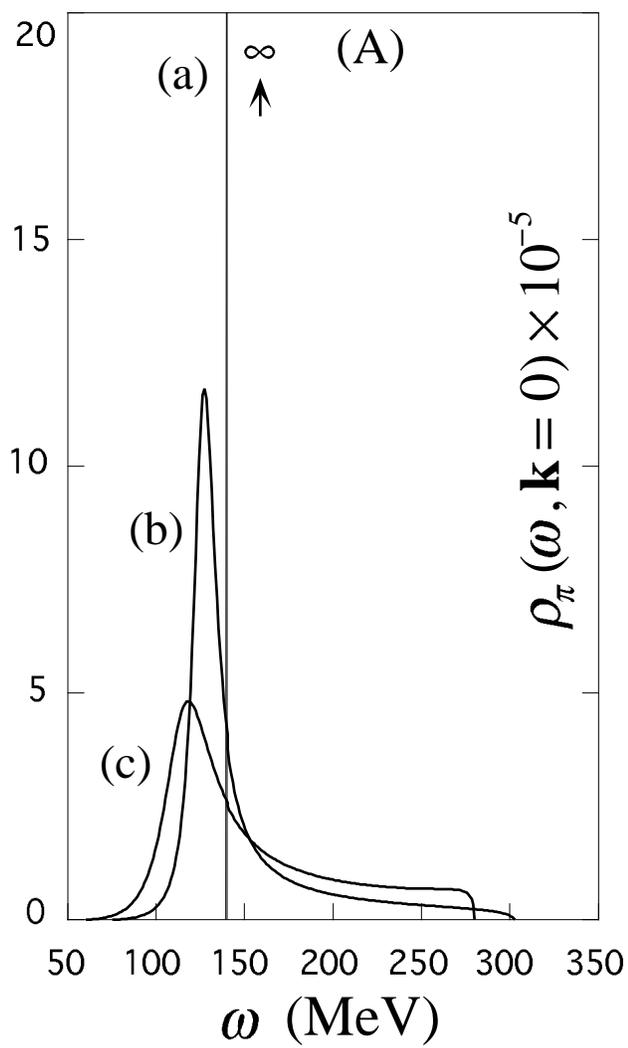
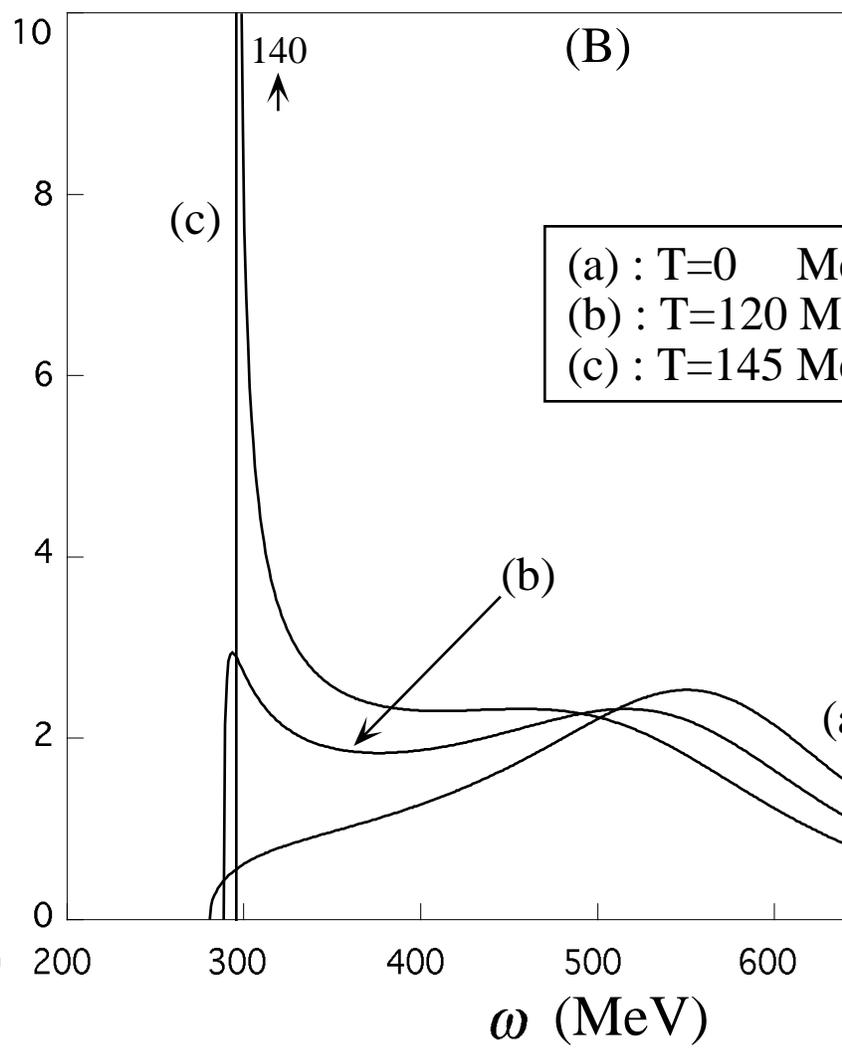
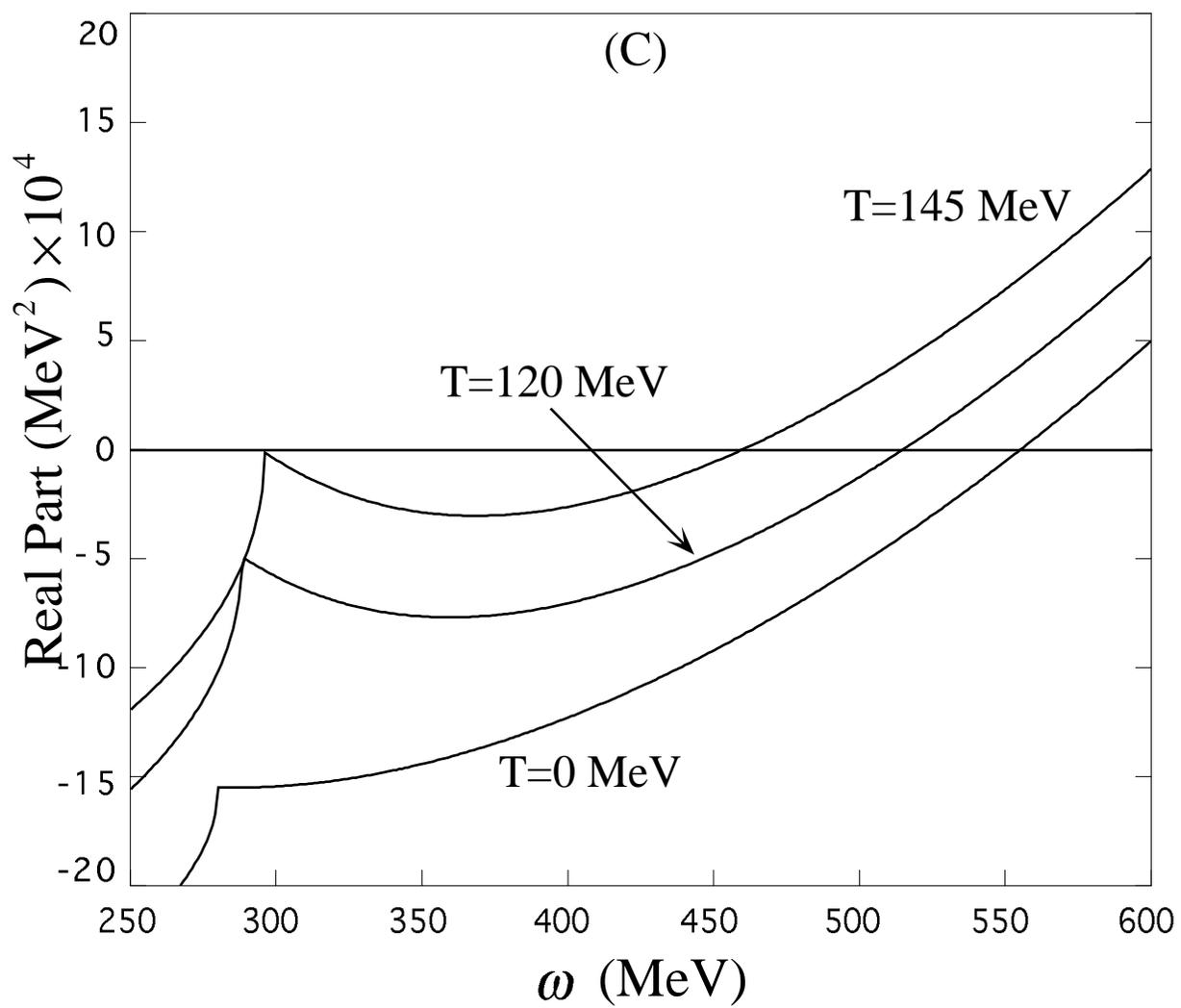
\end{document}